# Mystery of superconductivity in FeTe films and the role of neighboring layers

Xiong Yao,[1,2,a)] Hee Taek Yi,[3] Deepti Jain,[3] Xiaoyu Yuan,[3] and Seongshik Oh[2,a)]

[1]*Ningbo Institute of Materials Technology and Engineering, Chinese Academy of Sciences, Ningbo 315201, China*

[2]*Center for Quantum Materials Synthesis and Department of Physics & Astronomy, Rutgers, The State University of New Jersey, Piscataway, New Jersey 08854, United States*

[3]*Department of Physics & Astronomy, Rutgers, The State University of New Jersey, Piscataway, New Jersey 08854, United States*

Since the discovery of superconductivity in the Fe(Te,Se) system, it has been a general consensus that the end member of FeTe is not superconducting. Nonetheless, in recent years, there have been reports of superconducting FeTe films, but the origin of their superconductivity remains mysterious. Here, we provide the first comprehensive review of all the reported FeTe films regarding the relationship between their superconductivity and neighboring layers. Based on this review, we show that telluride neighboring layers are the key to superconducting FeTe films. Then, with additional new studies, we show that stoichiometric Te content, which can be readily achieved in FeTe films with the assistance of neighboring telluride layers, might be crucial to stabilizing the superconductivity in this system. This work provides insights into the underlying mechanism behind superconductivity in FeTe films and sheds light on the critical role of neighboring layers and stoichiometry control toward manipulating topological superconductivity in FeTe heterostructures.

Recently Fe(Te,Se) system has been attracting intensive interest since signatures for topological superconductivity and Majorana modes were observed in the bulk compounds of Fe(Te,Se) by scanning tunneling microscopy (STM) and angle-resolved photoelectron spectroscopy (ARPES)[1-5]. Compared to other topological superconductor candidates[6,7], Fe(Te,Se) shows several advantages such as a relatively high $T_C$ above 10 K and robust superconductivity against aging and disorder[8]. A recent study shows that in Fe(Te,Se) bulk crystals, superconductivity is observed only if Fe/(Te+Se) stoichiometry is slightly below one, and topological surface state occurs only if Te/Se ratio is beyond a critical value[9]. These observations indicate that precise compositional control is critical for the topological and superconducting

---

[a)] Author to whom correspondence should be addressed.  Electronic mail: yaoxiong@nimte.ac.cn; ohsean@physics.rutgers.edu

properties in Fe(Te,Se) superconductors. However, it has long been a consensus that superconductivity is absent in pure FeTe system, and instead FeTe turns into an antiferromagnetic (AFM) correlated metal at low temperatures[10-14].

There are reports that superconductivity can emerge in oxygenated FeTe films[15-17], where excess Fe is believed to be oxidized, as similar effects are also observed in an annealing study of Fe(Te,Se) bulk crystals[18]. Without oxygen incorporation, pure FeTe films have been known to be non-superconducting (non-SC), according to multiple reports[13,14,17,19-21]. However, it was recently found that superconductivity can emerge in almost pure FeTe films with a minuscule level of unintentional Se impurities (Se/Te < 0.03) when grown on $Bi_2Te_3$ or MnTe layers[22,23]: without the neighboring $Bi_2Te_3$ or MnTe layers, it is known that at least ~10% of Se is needed to make Fe(Te,Se) superconducting[8,10,12,16,24]. Moreover, superconductivity with a sharp transition has been observed in FeTe films when interfaced with various telluride layers[19,20,25-28]. The origin behind these conflicting reports of the presence or absence of superconductivity in FeTe films remains unresolved. Here, we conduct a comprehensive review on the relationship between superconductivity and neighboring layers in FeTe films, with the aim of providing key insights into the underlying origin of the superconductivity.

**REVIEW ON FeTe HETEROSTRUCTURES AND THEIR NEIGHBORING LAYERS**

We summarized all the reported FeTe heterostructure films in Table 1, together with the characteristics of their neighboring layers: here, and in the rest of the paper, the notation, A/B, implies "A" grown on top of "B". These films are divided into SC and non-SC regime, by different background colors in the table. The superconducting transition temperatures of these films are defined by $T_C$-onset, $T_C$-50% $R_n$ (the temperature at which the resistance decreases to 50% of its normal state value) and $T_C$-zero, respectively. We classified the neighboring layers into four categories based on their properties, as listed in Table 1.

The first column of the neighboring layers in Table 1 is about the role of their in-plane crystalline symmetry with respect to superconductivity in FeTe thin films. Considering that lattice matching is critical to achieving high quality epitaxial films for most compound materials, 4-fold substrates like $SrTiO_3$ and MgO are the common choice of substrates for FeTe films, which also have 4-fold in-plane symmetry, and in these films, superconductivity is absent as expected[13,14,17]. Similarly, pure FeTe films interfaced with (4-fold) ZnSe layers, either in the stacking sequence of FeTe/ZnSe or ZnSe/FeTe, belong to the non-SC regime[19], as shown in Figure 1d. However, Figure 1c presents a recent study on FeTe film grown on 4-fold CdTe substrate, demonstrating a clear superconducting transition at 9.0 K, despite the absence of zero resistance down to 2 K[29]. Recently, our group discovered a novel "hybrid symmetry epitaxy" mode that facilitates high-quality epitaxial growth of 4-fold FeTe on top of 6-fold $Bi_2Te_3$ or MnTe, both demonstrating distinct signatures of superconductivity (Figure 1a and b)[22,23]. In the absence of a $Bi_2Te_3$ or MnTe buffer layer, FeTe film directly grown on 6-fold $Al_2O_3$ does not exhibit superconductivity[21]. On the other hand, it has been reported that superconductivity arises when $Bi_2Te_3$ or $Sb_2Te_3$ is deposited onto otherwise non-SC FeTe[19,20], as shown in Figure 1d. Along a similar line, Chang et al. observed the emergence of superconductivity when magnetic topological insulator $(Cr,Bi,Sb)_2Te_3$ or $MnBi_2Te_4$ is grown on FeTe[25,27,28]. Moreover, they also observed a sharp superconducting transition when a 6-fold ferromagnetic $CrTe_2$ layer is deposited on FeTe[26]. To summarize, although most of the SC FeTe films have 6-fold-symmetric neighboring layers, two counter-examples – SC FeTe film on 4-fold CdTe and non-SC FeTe film on 6-fold $A_2O_3$ – imply that the symmetry of the neighboring layers is not a critical factor behind the emergence of superconductivity in FeTe films.

The second column of Table 1 is about the conductivity of the neighboring layers. All the FeTe films grown on insulating oxide substrates (SrTiO$_3$, MgO, Al$_2$O$_3$), as well as the ZnSe buffer layer, are non-SC, according to Table 1. On the other hand, although all the FeTe heterostructures with conducting neighboring layers in Table 1 are SC, those on insulating CdTe substrate or MnTe layer also show either partial (CdTe) or full (MnTe) superconducting transition[23,29]. This shows that a conducting neighboring layer is not a prerequisite for superconductivity in FeTe films.

The third column of Table 1 is about the role of magnetism of the neighboring layers. The neighboring layers include non-magnetic, antiferromagnetic, and ferromagnetic materials. With a non-magnetic neighboring layer, FeTe films can be either superconducting or non-superconducting in Table 1. On the other hand, all the FeTe films in Table 1 with a magnetic neighboring layer are superconducting. Typically, ferromagnetic order is believed to compete with superconductivity. However, recent reports suggest the coexistence of ferromagnetism and superconductivity in (Cr,Bi,Sb)$_2$Te$_3$/FeTe (Figure 2a) and CrTe$_2$/FeTe heterostructures[25,26], where the large upper critical field is believed to protect the superconductivity against the pair-breaking effect induced by ferromagnetic exchange coupling. Moreover, a clear ferromagnetic hysteresis loop was observed in the Hall resistance measurement in these heterostructures at normal state (above the superconducting transition temperature). The FeTe films interfaced with antiferromagnetic MnTe (Figure 1b) or MnBi$_2$Te$_4$ (Figure 2b) layers also exhibit clear superconductivity. The magnetic order in pure FeTe bulk crystals was determined to be antiferromagnetic by neutron diffraction study[30], and the occurrence of bulk superconductivity in Fe(Te,Se) was attributed to the suppression of magnetic correlations achieved through Se doping according to a previous study[12]. It has been theoretically predicted that coherence–incoherence crossover in Fe-based superconductors is driven by the Hund's coupling[31]. In particular, FeTe shows a peculiar double-stripe magnetic configuration, in contrast to the single-stripe states in the parent compounds of other Fe-based superconductors[32]. Accordingly, whether neighboring magnetic orders can suppress the underlying AFM order of FeTe and trigger superconductivity or not is an insightful and unexplored question. Nonetheless, considering that superconducting FeTe states are observed for both magnetic and non-magnetic neighboring layers, magnetism is not a key factor behind the emergence of superconductivity in FeTe.

The last column in Table 1 categorizes these FeTe heterostructures based on whether their neighboring layers are tellurides or not. We observe a clear correlation between the presence of superconductivity in FeTe heterostructures and telluride neighboring layers. Specifically, all the films with telluride neighboring layers exhibit some form of superconductivity, while there have been no reports of superconducting FeTe films with non-telluride neighboring layers so far. The first superconducting FeTe heterostructure was discovered when $Bi_2Te_3$ is deposited on top of FeTe[19] (i.e. $Bi_2Te_3$/FeTe), where $Bi_2Te_3$ is a topological insulator (TI). In recent years, more variant versions of this heterostructure have been successfully synthesized, such as the inverted FeTe/$Bi_2Te_3$ configuration, as well as the magnetic variants $(Cr,Bi,Sb)_2Te_3$/FeTe and $MnBi_2Te_4$/FeTe[22,25,28,29]. Notably, superconductivity in the original $Bi_2Te_3$/FeTe and $Sb_2Te_3$/FeTe heterostructures is reported to be of an interfacial nature, as indicated by upper critical field measurements[19,20]. In contrast, the superconductivity in the other three heterostructures (FeTe/$Bi_2Te_3$, $(Cr,Bi,Sb)_2Te_3$/FeTe and $MnBi_2Te_4$/FeTe) appears to be bulk-like, as evidenced by an isotropic upper critical field[22,25,28]. The isotropic upper critical field suggests that the emergent superconductivity originates from the bulk of the FeTe layer. However, the effective superconducting thickness seems to be less than the FeTe thickness in $(Cr,Bi,Sb)_2Te_3$/FeTe and $MnBi_2Te_4$/FeTe, indicating that the superconductivity in these heterostructures may arise preferentially near the interfaces. Additionally, several other FeTe heterostructures based on non-TI telluride layers also demonstrate evidence of superconductivity, including FeTe/MnTe, FeTe/CdTe and CrTe$_2$/FeTe[23,26,29], and this implies that the topological surface states are not a critical factor for the emergence of superconducting FeTe state, either.

**ADDITIONAL STUDIES ON FeTe$_x$/Bi$_2$Te$_3$**

According to all the above observations, the only common factor among the superconducting FeTe films is that they are neighbored by telluride layers. Then the question is: what is the role of neighboring telluride layers in inducing superconductivity in FeTe thin films? One possibility is that the presence of telluride neighboring layers facilitates superconductivity in FeTe by stabilizing the Te stoichiometry with the assistance of the surrounding Te environment. In order to address this question, we grew two kinds of FeTe films in identical conditions using molecular beam epitaxy (MBE), one on a telluride bottom layer and the other on a non-telluride layer: FeTe$_x$/Bi$_2$Te$_3$ and FeTe$_x$/CaF$_2$. The exact composition of all our samples is determined by Rutherford backscattering spectroscopy (RBS). All the FeTe$_x$ films discussed below, grown in our MBE system, contain a minuscule level of unintentional Se impurities (Se/Te ≈ 1~3%). We have previously reported superconducting FeTe/Bi$_2$Te$_3$ films with optimized Te stoichiometry (FeTe$_{0.97}$Se$_{0.03}$), exhibiting a sharp superconducting transition with zero resistance temperature above 10 K[22]. Here we grew three FeTe$_x$/Bi$_2$Te$_3$ samples with varied Te stoichiometry, following the protocols in our previous report. Figure 3 shows the transport properties of these samples together with the corresponding reflection high-energy electron diffraction (RHEED) patterns. Superconductivity is observed in the sample of FeTe$_{1.1}$/Bi$_2$Te$_3$ (Figure 3a), while no signature of superconducting transition has been observed in FeTe$_{0.94}$/Bi$_2$Te$_3$ (Figure 3b) and FeTe$_{0.88}$/Bi$_2$Te$_3$ (Figure 3c). These observations unambiguously demonstrate a correlation of superconductivity and Te stoichiometry in FeTe$_x$/Bi$_2$Te$_3$. The ideal Te stoichiometry close to 1 lead to optimized superconductivity. Even though deviation from the ideal Te stoichiometry (both Te > 1 and Te < 1) causes suppression of superconductivity, the Te-poor condition leads to a more pronounced suppression effect than the Te-rich condition. Interestingly, the RHEED patterns of Te-poor samples are sharper and cleaner than those of Te-rich samples, as shown in Figure 3. These observations imply that Te stoichiometry is more critical than structural quality (RHEED pattern) for observing superconductivity in FeTe$_x$/Bi$_2$Te$_3$.

**ADDITIONAL STUDIES ON FeTe$_x$/CaF$_2$**

Then we grew and investigated FeTe$_x$/CaF$_2$ films in a similar way. We chose CaF$_2$ substrate because the Fe(Te,Se) films grown on CaF$_2$ are known to exhibit the highest $T_C$[10]. We grew the FeTe$_x$ films with a thickness of around 12 nm on 10 × 10 × 0.5 mm$^3$ CaF$_2$(001) substrates. CaF$_2$ substrates were cleaned ex situ by 5 minutes of exposure to UV-generated ozone and in situ by heating to 400 °C for 10 minutes. The growth conditions are similar to that of FeTe$_x$/Bi$_2$Te$_3$. Figure 4a presents the temperature-dependent sheet resistance of FeTe$_{0.87}$, in which the superconducting transition is absent, and instead, an upturn behavior appears at low temperatures. It also exhibits the characteristic hump feature around 70 K, which is believed to be related to the development of the AFM order[33]. Overall, the R vs T behavior is similar to those of Fe(Te,Se) films grown on MnTe under a Te-poor environment[23]. The RHEED patterns exhibit bright streaks with no irregularities on the pattern, implying that the film is structurally pristine, despite the presence of excess Fe.

Since Te flux is insufficient in Figure 4a, we increased the Te flux and grew another sample in Figure 4b. The composition is determined as $FeTe_{1.15}$ based on RBS. The RHEED patterns exhibit ill-defined streaks with prominent spots on the streaks, suggesting that the structural quality of the film deteriorates as Te flux gets higher. Figure 4b gives the temperature-dependent sheet resistance of $FeTe_{1.15}/CaF_2$: interestingly, even though the resistance does not fully reach zero, a clear superconducting transition can still be observed at around 10 K. This suggests that this sample has some fraction of superconducting phase[13,23,34]. Both $FeTe_x/Bi_2Te_3$ and $FeTe_x/CaF_2$ samples indicate that the Te-rich condition favors superconductivity, whereas Te-poor samples do not exhibit any hints of superconductivity despite the better structural quality. The fact that the superconducting FeTe phase, albeit incomplete, can be achieved even without neighboring telluride layers suggests that superconductivity is likely to be an intrinsic part of the $FeTe_x$ phase diagram within some small compositional window.

**CONCLUSIONS**

We tried multiple approaches to improve the Te stoichiometry in $FeTe_x/CaF_2$ films, such as annealing the Te poor samples with Te flux supplied. Still, all these attempts failed to achieve 1:1 stoichiometry, lacking any hints of superconductivity. This is in stark contrast to the FeTe films grown on $Bi_2Te_3$ or MnTe[22,23], in which superconductivity can be readily stabilized over a wide growth window. The above results suggest that the neighboring telluride layers facilitate the stability of the Te stoichiometry, favoring the emergence of the superconducting phase. In other words, neighboring with stoichiometrically stable telluride compounds such as $Bi_2Te_3$ ($Sb_2Te_3$), MnTe, CdTe, and $CrTe_2$ seems to help stabilize the stoichiometry of FeTe close to 1:1 while suppressing excess Fe. Superconductivity in FeTe films seem to be more vulnerable to excess Fe than excess Te, as indicated by the fact that superconductivity can survive in both $FeTe_x/Bi_2Te_3$ and $FeTe_x/CaF_2$ with excess Te up to 10~15%. The telluride neighboring layer might play a role as a reservoir layer for Te atoms, rebalancing the Te stoichiometry of the FeTe layer by injecting or withdrawing Te atoms through interdiffusion. Such a mechanism explains the wider growth window for superconducting FeTe films on telluride neighboring layers. The optimized Te stoichiometry by telluride neighboring layers might also play a key role in the emerging superconductivity in $Bi_2Te_3/FeTe$ and $Sb_2Te_3/FeTe$ heterostructures, where topological surface states of the topological insulators ($Bi_2Te_3$ and $Sb_2Te_3$) were believed to be the driving mechanism behind the unexpected superconductivity[19,20,25,27,28].

In summary, we reviewed all the existing FeTe films while focusing on the relationship between the superconductivity and their neighboring layers. From this review, we showed that the presence of neighboring telluride layers is the only common factor behind the superconducting FeTe films. Nonetheless, the role of telluride neighboring layers for the superconductivity remains elusive. By comparing $FeTe_x/Bi_2Te_3$ and $FeTe_x/CaF_2$ films grown in a range of stoichiometric conditions, we revealed that the emerging superconductivity in $FeTe_x$ films is sensitive to the Te stoichiometry. In particular, while the growth window for the superconducting phase is broad for $FeTex/Bi_2Te_3$ films, the growth window for the superconducting phase is extremely narrow for $FeTe_x/CaF_2$ films. Also, we observed that excess Te, rather than excess Fe, is more favorable for superconductivity in $FeTe_x$ films. Combining all these observations, we propose that the neighboring telluride layers facilitate the stability of the Te stoichiometry of FeTe and favor the emergence of the superconducting phase. Even though this mechanism still remains a hypothesis and requires further in-depth investigation, our study provides unique insights into the origin of superconductivity in FeTe thin films and the role of neighboring layers, shedding light on how to manipulate topological and superconducting properties in FeTe heterostructures.

| Heterostructure | | | | Neighboring Layer | | | | |
|---|---|---|---|---|---|---|---|---|
| SC vs Non-SC | $T_C$-onset | $T_C$-50%$R_n$ | $T_C$-zero | 4-fold vs 6-fold | Conducting vs Insulating | Magnetic vs Non-magnetic | Telluride vs Non-Telluride | Ref. |
| FeTe/Bi$_2$Te$_3$ | 12.9 K | 12.1 K | 10.7 K | 6-fold | Conducting | Non-magnetic | Telluride | 22 |
| FeTe/MnTe | 11.0 K | 9.7 K | 4.1 K | 6-fold | Insulating | Magnetic (AFM) | Telluride | 23 |
| FeTe/CdTe | 9.0 K | 2.8 K | - | 4-fold | Insulating | Non-magnetic | Telluride | 28 |
| FeTe/ZnSe (ZnSe/FeTe) | | | | 4-fold | Insulating | Non-magnetic | Non-telluride | 19 |
| FeTe/SrTiO$_3$ | | | | 4-fold | Insulating | Non-magnetic | Non-telluride | 13,14 |
| FeTe/MgO | | | | 4-fold | Insulating | Non-magnetic | Non-telluride | 17 |
| FeTe/Al$_2$O$_3$ | | | | 6-fold | Insulating | Non-magnetic | Non-telluride | 21 |
| Bi$_2$Te$_3$/FeTe | 12.0 K | 10.6 K | 10.0 K | 6-fold | Conducting | Non-magnetic | Telluride | 19 |
| Sb$_2$Te$_3$/FeTe | 12.3 K | 6.3 K | 3.1 K | 6-fold | Conducting | Non-magnetic | Telluride | 20 |
| (Cr,Bi,Sb)$_2$Te$_3$/FeTe | 12.4 K | 11.9 K | 11.4 K | 6-fold | Conducting | Magnetic (FM) | Telluride | 24 |
| MnBi$_2$Te$_4$/FeTe | 12.1 K | 11.5 K | 10.8 K | 6-fold | Conducting | Magnetic (AFM) | Telluride | 27 |
| CrTe$_2$/FeTe | 12.7 K | 11.9 K | 11.0 K | 6-fold | Conducting | Magnetic (FM) | Telluride | 25 |

Table 1. Summary of all the existing FeTe heterostructures and the corresponding neighboring layers in the context of emerging superconductivity.

FIGURES

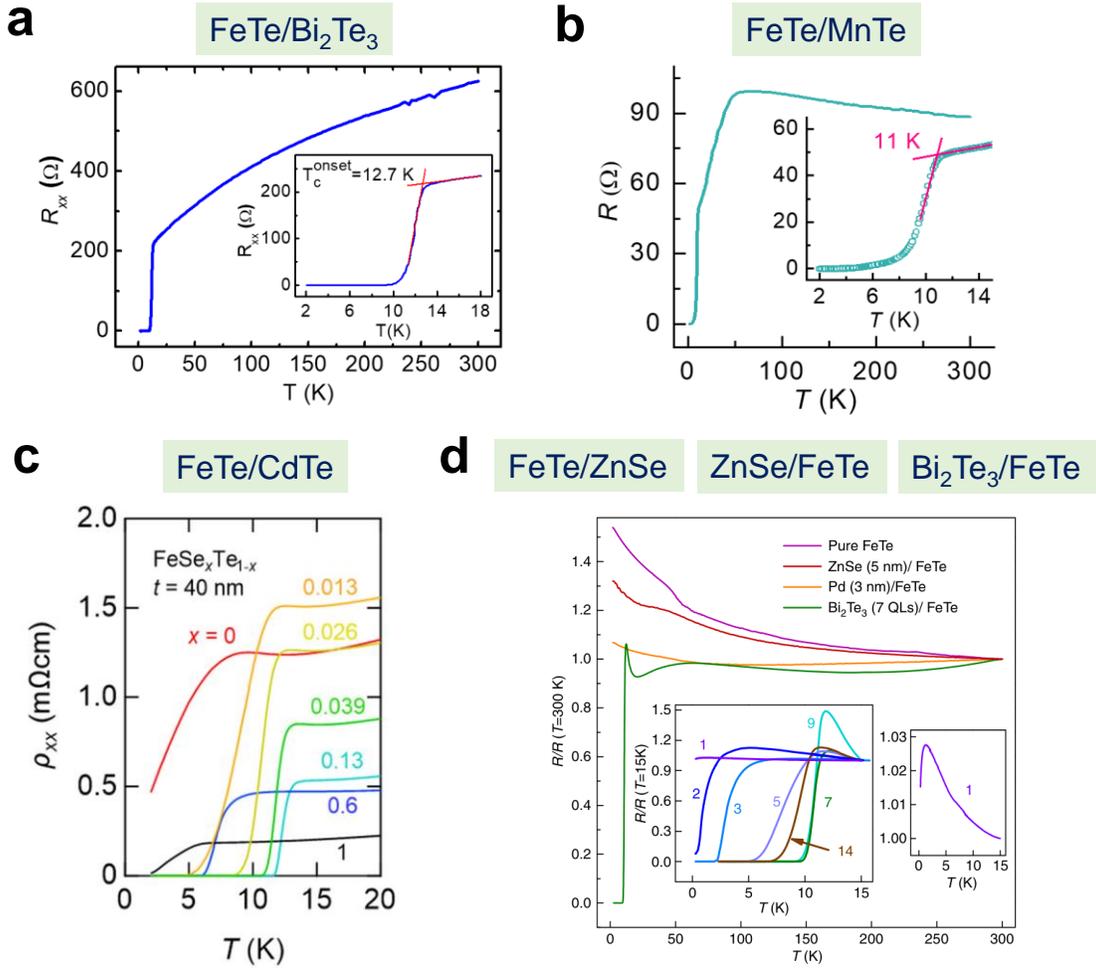

FIG. 1. Superconducting FeTe heterostructures with telluride neighboring layers. (a) Temperature-dependent longitudinal resistance of a FeTe(with ~3% unintentional Se)/Bi$_2$Te$_3$ film from 300 to 2 K, adapted from ref. [22] (b) Temperature-dependent longitudinal resistance of an FeTe (22 nm, with ~3% Se)/MnTe (10 nm) film from 300 to 2 K, adapted from ref. [23]. (c) Resistivity vs Temperature ($\rho_{xx}$-T) curves for FeSe$_x$Te$_{1-x}$ thin films grown on CdTe substrate with different Se content x, adapted from ref.[29]. (d) Temperature-dependent normalized resistance of pure FeTe(on ZnSe buffer), ZnSe/FeTe, and Pd/FeTe films, as well as a Bi$_2$Te$_3$/FeTe heterostructure, adapted from ref. [19]. The bottom left inset shows the normalized resistance versus temperature of Bi$_2$Te$_3$/FeTe heterostructures with varied Bi$_2$Te$_3$ thicknesses, while the bottom right inset is an enlarged plot for a Bi$_2$Te$_3$(1 QL)/FeTe heterostructure.

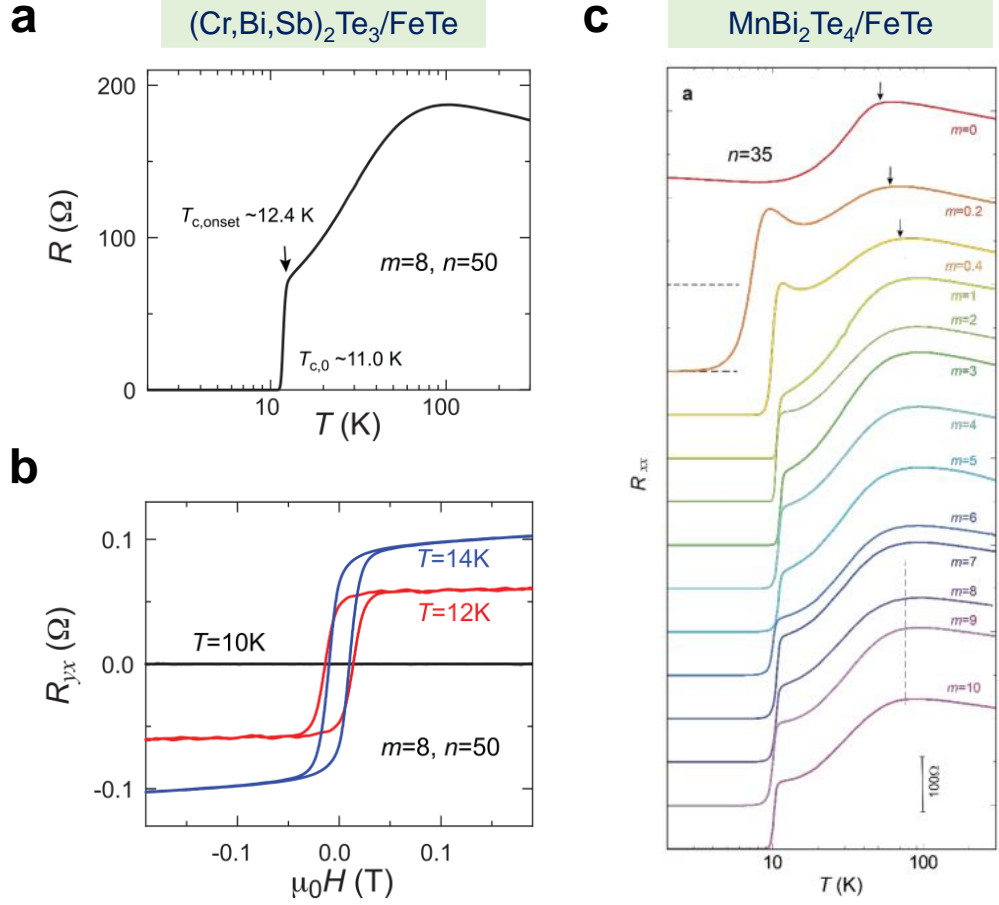

FIG. 2. Superconducting FeTe heterostructures with magnetic topological insulators as neighboring layers. (a) Temperature-dependent sheet longitudinal resistance and (b) Magnetic field-dependent Hall resistance $R_{yx}$ at varied temperatures of Cr-doped $(Bi,Sb)_2Te_3$/FeTe heterostructures, adapted from ref. [25]. (c) Temperature dependence of the sheet longitudinal resistance $R_{xx}$ of the $MnBi_2Te_4$/FeTe heterostructures. Here m refers to the thickness of the $MnBi_2Te_4$ layer, and n refers to the thickness of the FeTe layer. The two horizontal dashed lines represent zero resistance in the m = 0 and m = 0.2 heterostructures, adapted from ref. [28].

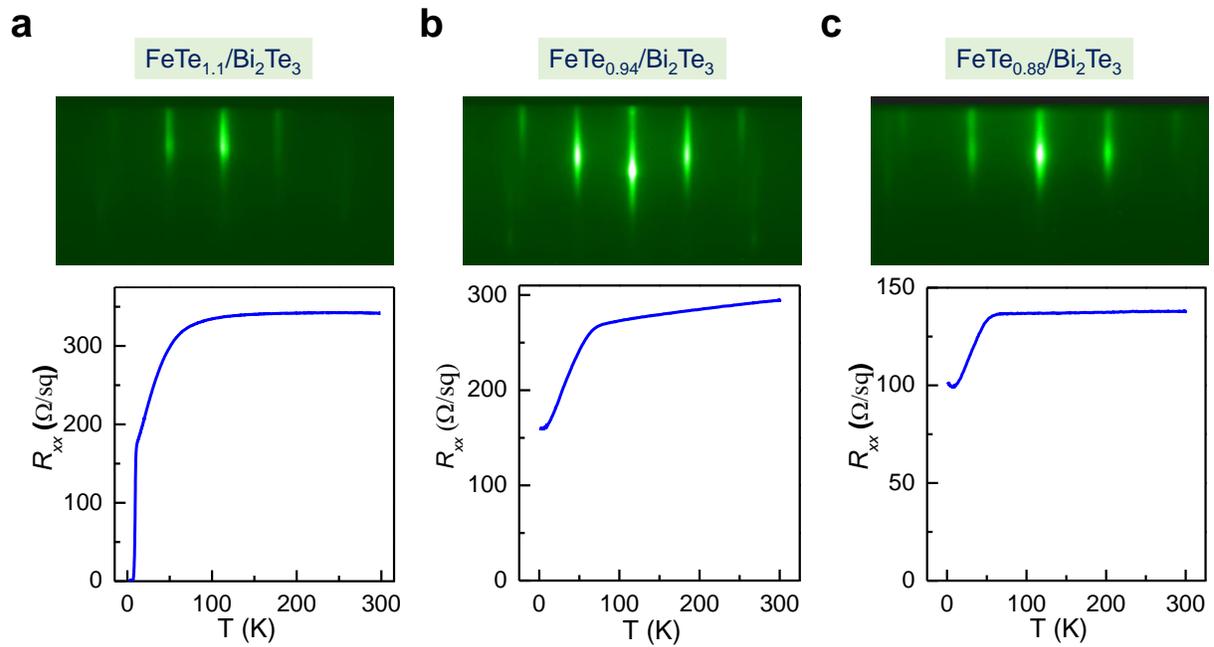

FIG. 3. Temperature-dependent longitudinal sheet resistance and RHEED patterns of three FeTe$_x$/Bi$_2$Te$_3$ samples with varied Te stoichiometry of (a) x = 1.1, (b) x = 0.94 and (c) x = 0.88.

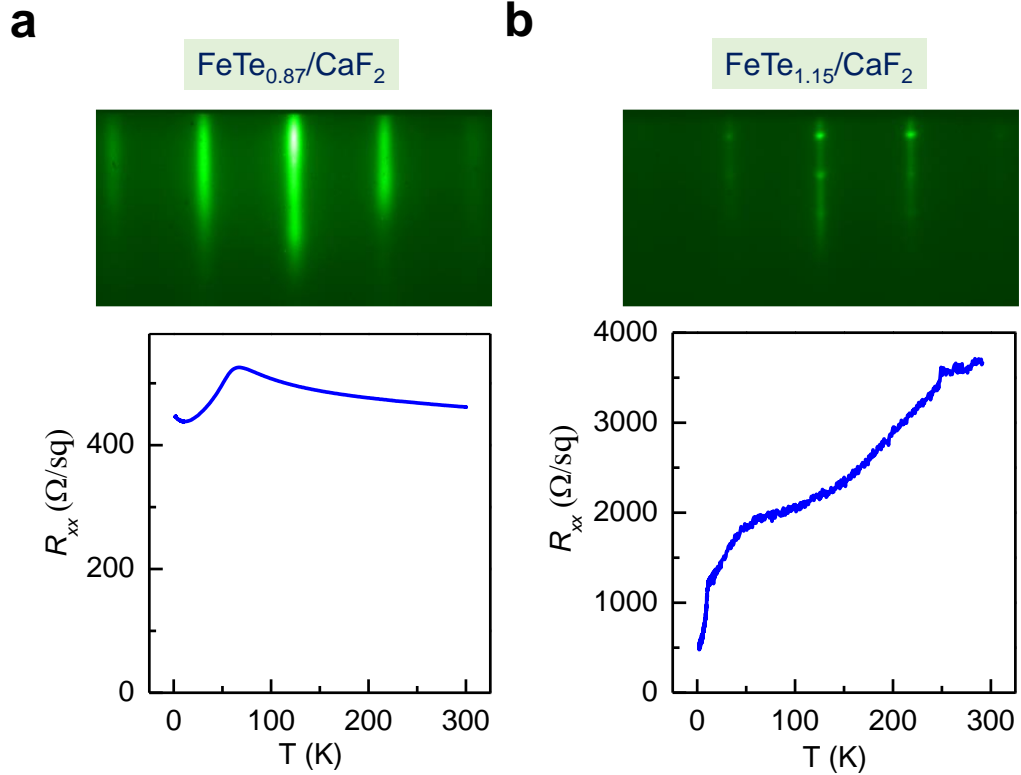

FIG. 4. (a) Temperature-dependent longitudinal sheet resistance and the corresponding RHEED pattern of a 12 nm FeTe$_{0.87}$/CaF$_2$ from 300 to 2 K. (b) Temperature-dependent longitudinal sheet resistance and the corresponding RHEED pattern of a 12 nm FeTe$_{1.15}$/CaF$_2$ from 300 to 2 K: note the presence of incomplete superconducting transition at 11 K.

## ACKNOWLEDGMENTS

The work at Rutgers is supported by Army Research Office's W911NF2010108, MURI W911NF2020166, and the center for Quantum Materials Synthesis (cQMS), funded by the Gordon and Betty Moore Foundation's EPiQS initiative through grant GBMF10104. X. Yao is supported by the National Natural Science Foundation of China (Grant No. 12304541).